# Implementing an Edge-Fog-Cloud architecture for stream data management

Lilian Hernandez
*People in Motion Lab*
*University of New Brunswick*
*Fredericton, NB, Canada*
lhernand@unb.ca

Hung Cao
*People in Motion Lab*
*University of New Brunswick*
*Fredericton, NB, Canada*
hcao3@unb.ca

Monica Wachowicz
*People in Motion Lab*
*University of New Brunswick*
*Fredericton, NB, Canada*
monicaw@unb.ca

*Abstract—* The Internet of Moving Things (IoMT) requires support for a data life cycle process ranging from sorting, cleaning and monitoring data streams to more complex tasks such as querying, aggregation, and analytics. Current solutions for stream data management in IoMT have been focused on partial aspects of a data life cycle process, with special emphasis on sensor networks. This paper aims to address this problem by developing streaming data life cycle process that incorporates an edge/fog/cloud architecture that is needed for handling heterogeneous, streaming and geographically-dispersed IoMT devices. We propose a 3-tier architecture to support an instant intra-layer communication that establishes a stream data flow in real-time to respond to immediate data life cycle tasks in the system. Communication and process are thus the defining factors in the design of our stream data management solution for IoMT. We describe and evaluate our prototype implementation using real-time transit data feeds. Preliminary results are showing the advantages of running data life cycle tasks for reducing the volume of data streams that are redundant and should not be transported to the cloud.

## I. INTRODUCTION

One of the main concerns in the era of the Internet of Moving Things (IoMT) is the risk of overflowing a system due to billions of IoMT devices generating a huge volume of data streams that need to be sent out to the cloud for processing and analytic tasks. Recent studies [1–7] have demonstrated the importance of combining edge and cloud computing in stream data management to address the issues of speed of execution, accuracy, bandwidth cost and privacy. In contrast, [2] points out that fog computing should also be considered as an extension (not a replacement) of cloud computing mainly because fog computing can run processing and analytics that clean and aggregate the data streams before sending them up to the cloud. Some experiments of combining fog and cloud computing in smart cities [8], smart factories [9], and dairy farming [10] are showing the optimization of streaming workflows and cost-minimization for stream big data processing in geographically distributed datacenters.

This paper proposes a 3-tier architecture for combining edge, fog and cloud computing that is needed to provide the means for a data life cycle over transit data feeds that can reinforce a data flow from things to the cloud, passing through edge and fog nodes. To the best of our knowledge, an edge-fog-cloud architecture has not yet been proposed in the research literature. Moreover, very few is known on suitable data life cycle approaches for IoMT [17].

Our research challenge is two-fold:

- how to handle the complexity of the data life cycle not only because of the increasing data rates, but also because of the need for adopting an efficient and transparent exchange of data between edge nodes and fog nodes that will allow numerous feedback loops and re-running of workflow tasks;

- how to automate and improve workflow tasks performed on IoMT data streams (e.g. control flow, monitoring, and task sequence) in conjunction with computational tasks on the same data streams (e.g. capture, querying, pre-processing). Currently, data life cycle approaches are based on sequences of tasks that are programmed independently, making them unsuitable for IoMT.

The main scientific contributions of our paper are:

- development of an end-to-end architecture by combining edge, fog and cloud computing for IoMT data-intense applications;

- development of a data life cycle approach to capture the dynamicity of IoMT data, i.e. the fact that they are produced incrementally, regenerated, modified or temporarily unavailable.

Our objective is to provide a formal 3-tier architecture to facilitate IoMT data flows based on an agnostic execution model which enables data life cycle management.

The remaining of this paper is organized as follows. Section 2 introduces the concept of streaming data life cycle and propose a formal model that allows to expose an end-to-end life cycle across heterogeuous architecture levels. Section 3 describes the 3-tier architecture. Its implementation and the preliminary results of an experiment using transit data feeds are described in Section 3. Section 4 concludes the paper by sharing our future research work.

## II. STREAMING DATA CYCLE MODEL

The approach we propose follows the inherit goal of data life cycles which is to integrate the data flow from things, to the edge nodes, to the fog nodes, and finally to the cloud using an execution model that allows code execution of each workflow task. On the one hand, the execution model allows to describe a task sequence and data dependency such as explicit/implicit control flow in real-time or running continuous queries on IoT data streams. On the other hand, once the data life cycle is known and formally defined in the execution model, the workflow tasks are executed such as for automation of tiered storage; processing at any tier of the architecture; coordination between DSL links for IoMT data flows; monitoring task completion and data production; and so forth. Table I describes the main phases of our streaming data cycle model.

TABLE I. THE PROPOSED IoMT DATA CYCLE MODEL

| Phases | Objectives |
|---|---|
| Data Flow | DSL links |
| Execution Model | Task Sequence & Data Dependency |
| Control Flow | Explicit/Implicit. |
| Monitoring | Task Completion & Data Dependency |

The data streams enter an edge node after being acquired by an IoMT device, or created from some other data already are moved to the fog layer. Between these two points in time, the data progress through a series of different tasks of the workflow, such as data storage, data leverage, data acquisition, data control, etc. The tasks are not necessarily sequential since data does not have to pass through all the tasks. The 3-tier architecture is explained in the next section.

## III. SYSTEM ARCHITECTURE

We propose an end-to-end architecture based on the main characteristics of IoMT data streams as described by [16]:
- Each tuple in a stream arrives online.
- A system has no control over the order in which a tuple arrives within a data stream or across data streams.
- Data streams are potentially unbounded in size.

They consist of a sequence of out-of-order tuples containing attributes such as:

$$T1 = (S_1, x_1, y_1, t_1)$$

where

$S_1$: is a set of attributes containing information about each IoMT device.

$x_1, y_1, t_1$: is the geographical location of an IoMT device at the timestamp $t$.

### A. The 3-tier layer architecture

The overall architecture consists of the following layers: edge layer, fog layer, and cloud layer (Figure 1). The edge layer contains an edge node and it is in charge to acquire the tuples coming from the IoMT devices. The fog layer is formed by fog nodes and it is where the streaming data cycle model is executed. Finally, the cloud layer is where the data center is located. The communication among these three layers is performed by two principal components: the message broker and the distributed service links (DSL).

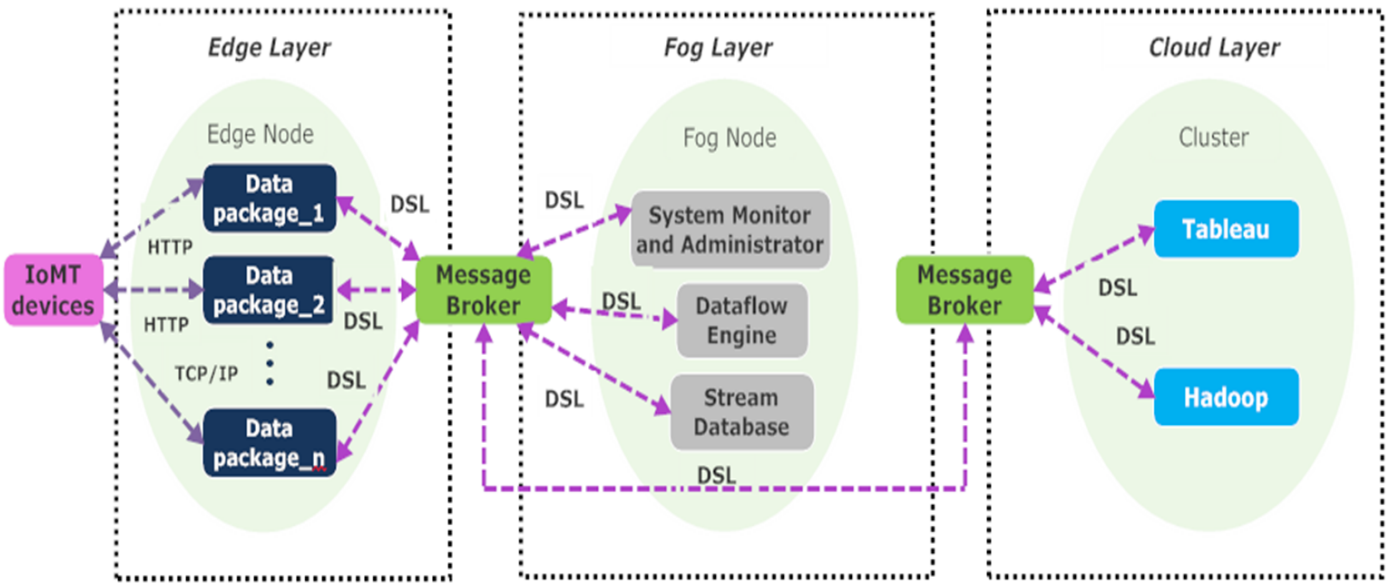

Fig. 1. Overview of our three-tier system architecture

The message broker decouples communication between the edge nodes and the fog nodes for invoking services to retrieve the data stream packages over a fixed time frequency, for example, every 5 minutes, 1 hour or a day. Moreover, it performs message aggregation, decomposing messages into multiple messages and sending them from the fog layer to the cloud. The message broker can also perform an instant intra-layer communication that allows them to establish data flows that can share data within the system. In this way the edge nodes handle the transportation of heterogeneous, streaming and geographically-dispersed IoMT data streams. In contrast, the fog nodes are designed to support a real-time data life cycle model.

*B. The streaming data workflow*

Six components are designed to support the workflow tasks of our data life cycle model. The components are: the system administrator, system monitor, data flow engine, message broker, communication links, and the stream database (Figure2). Based on these components, the system is enabled to treat IoMT data as data flows to communicate, process and transport data from end devices to the cloud.

Each task requires a specific data flow with unique purposes that can be either performing data acquisition, data processing, or data storage. One advantage of our proposed data life cycle process is the decision of having placed a fog node within a data stream management that can offer the possibility of starting the data life cycle process at anytime, instead of starting it only when the data streams have reached the cloud.

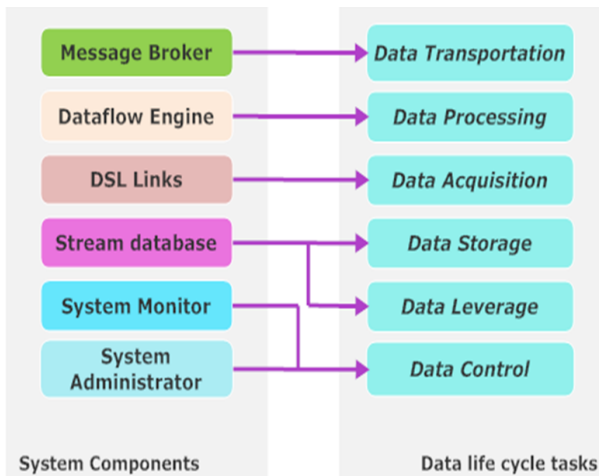

Fig. 2 Overview of the main tasks performed by the system components

The processing task is designed for removing errors and inconsistencies from the data streams stored in the stream database. Guaranteeing data quality for continuous and high volume of data streams is a nontrivial task, and performing this task automatically is even more challenging because the high data rate [11]. The Data Flow Engine is used to implement the processing task for handling (1) missing tuples, (2) duplicated tuples, (3) missing attribute values, (4) redundant attributes, and (5) wrong attribute values. Running the processing task at a fog node reduces the volume of data streams that need to be transported to the cloud.

The acquisition task at a fog node runs in a similar way as the MapReduce function in Hadoop. It sorts out the incoming tuples according to any attribute value of a tuple or a timestamp as soon as the tuples arrive at the fog node. It is important to mention that in our system architecture, the stream database is a fundamental to support the real-time data life cycle, whereas MapReduce is used to support the batch processing cycle in the cloud.

The cloud is an appropriate batch processing environment for large scale of data streams. An example of this use case is when it is required a strong resource to process a vast amount of data streams. Then, the cloud successfully provides the underlying infrastructure, platforms, and several services; such as pools of server, distributed storages, and networking resources, to distributed data processing among platforms. In addition, cloud computing models are supported to accelerate the potential of large-scale data processing functions that are needed to lead fully satisfied big data stream processing tasks. Therefore, the cloud computing environments are not limited to only offer flexibility and efficiency for accessing data streams; they also provide a high-performance computing power to analyze, and efficiently process data streams under cost-effective needs.

The real-time data life cycle takes place every time the tuples are sent from the IoMT devices to the edge node, and then they flow through the 3-tier architecture from the edge node to the cloud passing by the fog node as an intermediate node. The life cycle consists of six tasks defined as data transportation, data processing, data acquisition, data storage, data leverage, and data control. Figure 2 illustrates the relation between each task of the data life cycle, and the respectively system component within our system prototype.

The real-time data life cycle tasks and the way they work together with the system components to accomplish common functions can be described as follows: All tuples reach the edge node from the IoMT devices via HTTP or TCP/IP protocol. Every time a tuple arrives to the edge layer a set of tuples is formed as a sequence. Once the tuples are within the edge layer, they are sent (data transportation – message broker) to the fog layer by the edge node and received (data acquisition – DSL links) by the fog node. In order to process the tuples and perform preliminary analytics (data processing – data flow engine) the fog node controls the sets of tuples (data control – system monitor and administrator) by retaining (data storage – stream database) and retrieving (data leverage – stream database) the set of tuples continuously. Finally, the set of tuples that deserve to reach the cloud are retrieve (data leverage – stream database) and sent (data transportation – message broker) to the fog node to the cluster.

IV. SYSTEM IMPLEMENTATION AND RESULTS

For our experiment we have selected transit feed data from the CODIAC transit network for Greater Moncton area (Figure 3). The network consists of 21 bus routes operating from Monday to Saturday, some of which provide evening and

Sunday services. Every bus in the transit network has installed a GPS receiver for collecting its location every 5 seconds. The set of attributes in a tuple are listed in Table II.

TABLE II. TUPLE ATTRIBUTES

| Attribute | Description |
|---|---|
| 1. vlr_id | The ID of the data point in the vehicle location reports table. |
| 2. route_id_vlr | The route ID in the vehicle location reports table. |
| 3. route_name | The route name. |
| 4. route_id_rta | The route ID in the route transit authority table. |
| 5. route_nickname | The abbreviation of the route. |
| 6. trip_id_br | The trip ID in the bid route table. |
| 7. transit_authority_service_time_id | Transit authority service time ID. |
| 8. trip_id_tta | Transit authority trip ID. |
| 9. trip_start | Start time of the trip. |
| 10. trip_finish | Finish time of the trip. |
| 11. vehicle_id_vab | Vehicle ID. |
| 12. vehicle_id_vlr | Vehicle ID in the vehicle locations reports table. |
| 13. vehicle_id_vlr_ta | Descriptive name of the bus. |
| 14. bdescription | Bus description. |
| 15. lat | Latitude. |
| 16. lng | Longitude. |
| 17. timestamp | Timestamp of the data point. |

In this experiment, the edge node known as Cisco IR829 Industrial Integrated Services Router is envisaged to be installed in the future on the top of buses that form the tested transit system. It has an Intel Atom Processor C2308 (1M Cache, 1.25 GHz) Dual Core X86 64bit, 2GB DDR3 memory, 8MB SPI Bootflash, 8GB (4GB usable) eMMC bulk flash, and multimode 3G and 4G LTE wireless WAN and IEEE 802.11a/b/g/n WLAN connections. Because of it is resistant to shock, vibration, dust, humidity, and water spray, and a wide temperature range (-40°C to +60°C and type-tested at +85°C for 16 hours) [19], this type of router accomplishes the requirements of our system prototype. Besides, this edge node comes with two operating system: a Cisco IOS system that runs a standard Cisco IOS package which handles all the routing, switching, and networking; and a guest operating system IOx running on a virtual machine.

Regarding the fog node, we suggest that in the future implementation of our system prototype, one fog node will be installed per bus terminal. The proposed fog node distribution is shown on the map inset in Figure 3. Current commercial systems for deploying data life cycle models using fog nodes are presented in Table III. They are the Edge Fog Fabric (EFF) Cisco platform [13], the Segment platform [14] offered by Segment Company, the IBM Watson IoT platform for edge analytics [15], and the Axon Predict platform produced by Greenwave System [16]. We have selected the EFF Cisco platform for our implementation because this platform supports block-programming for fostering experimentation. By using this programming style, we are able to control the effects of a data life cycle model by observing the content of a data stream (input) and the results (outputs) of any task. However, more research is needed to evaluate other platforms in the future.

Figure 4 shows the dashboard of the EFF Cisco Platform running over Ubuntu 16.4 OS at the fog layer. The Server (fog node) specifications to fulfill are: to have free 4GB or more per CPU, to have previously installed the following package of libssl1.0.0 Version: 1.0.2g-1ubuntu4.6, and libicu55 Version: 55.1-7ubuntu0.1, and to obtain superuser (root) permissions or sudo access to complete the installation of the Cisco Platform.

Our execution model consists of four programming blocks as highlighted in Figure 4. The first block is implemented for performing the data acquisition task where the data stream packages are transformed into CSV files. The second block represents the creation of a unique identifier number (ID) for each tuple arriving from the pulled data streams. The third block is performing both data cleaning and data sorting tasks, as well as supporting local notifications such as alarm messages.

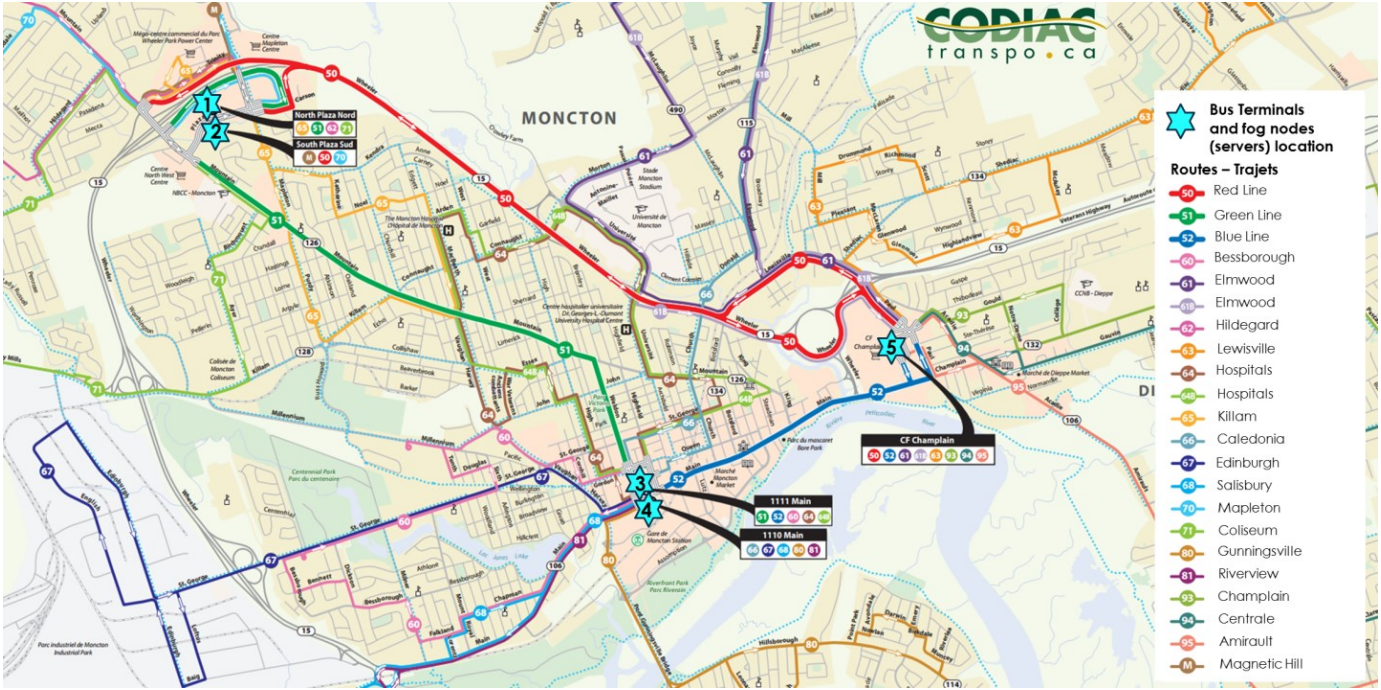

Fig. 3. Overview of the transit network used for the experiement.

The tuples were sorted by <route_id_rta, trip_id_br, timestamp>. We have implemented two alarm messages for missing tuples and duplicated tuples. Finally, the fourth block creates data tables. All data tables are stored temporally in the Stream database [18].

There were 65,097,658 tuples stored in the Stream database for a period of one year of streaming data from June $1^{st}$ 2016 to May $25^{th}$ 2017 from the edge nodes. After performing the data processing tasks at the fog layer, 38,653,787 tuples were deleted. The resulting data table consists of only 26,443,871 tuples which were then transported to the cloud. Table IV illustrates the statistics of one day (April $15^{th}$ 2017) from the one year experiement. In this case, 24,740 tuples have arrived in the cloud belonging to 16 bus routes that were running on that day.

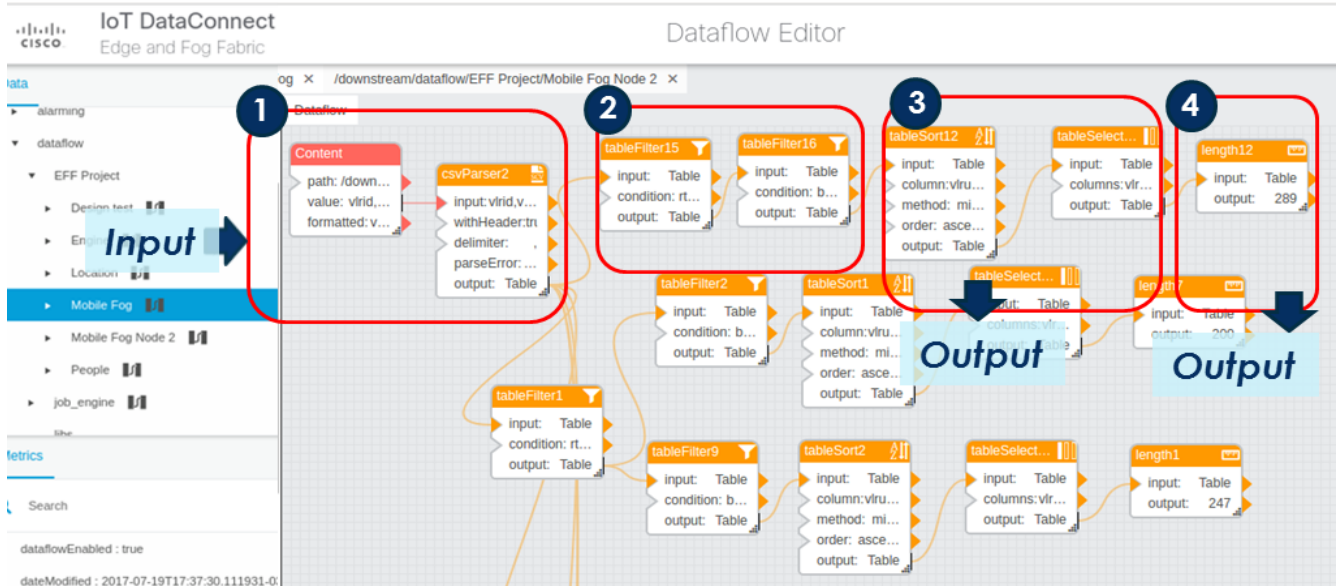

Fig. 4 Overview of the execution model

TABLE III. COMMERCIAL PLATFORMS FOR FOG COMPUTING

| Data life cycle tasks | EFF (Cisco) | Segment (Segment) | IBM Watson (IBM) | Axon Predict (Greenwave Systems) |
|---|---|---|---|---|
| Local Notification | Yes | Yes | Yes | Yes. |
| Processing | Yes. | No | Yes | Yes |
| Aquisition | DSL | Java API | Java API | Yes. The method is not specified. |
| Storage | Yes | Yes | Yes | Yes |
| Leverage | Yes | Yes | Yes | Yes |
| Data Control | Yes | Yes | Yes | Yes |

The tuples were sorted by <route_id_rta, trip_id_br, timestamp>. We have implemented two alarm messages for missing tuples and duplicated tuples. Finally, the fourth block

Although tuples from 478 trips were expected to arrive in the cloud, tuples from only 104 trips have actually arrived due to the data processing. These preliminanary results reveal the importance of fog computing in assuring the quality of data streams sent to the cloud, despite the drawback of deliverying a limited number of tuples for further analytics in the cloud.

In our proposed system, a Hadoop cluster includes a master node and a server node that will be deployed using the Compute Canada West Cloud. The large scale of data streams will be historically accumulated, and stored in the distributed file system of this cluster. It is worth to indicate that the Hadoop cloud resources are resilient and can be easily scaled up to afford the continuous growth of data streams on the cloud. Besides, it is important to mention that the high availability of the Hadoop cluster is preserved because the data streams are chunked into different partitions, and replicated through different nodes inside the cluster. The MapReduce programing model implemented in our cluster will handle the batch processing tasks in which tuples of data streams with the same key are mapped; whereas computing, and analyzing tasks are executed in the reduce phase in a parallel manner.

TABLE IV. OVERVIEW OF PRELIMINARY RESULTS

| Bus Route | Number of scheduled trips | Number of performed trips whose tuples reached the cloud | (%) |
|---|---|---|---|
| 50 | 31 | 2 | 6.45 |
| 51 | 65 | 6 | 9.23 |
| 52 | 65 | 5 | 7.69 |
| 60 | 31 | 2 | 6.45 |
| 61 | 32 | 19 | 59.38 |
| 62 | 31 | 19 | 61.29 |
| 63 | 32 | 3 | 9.38 |
| 64 | 32 | 19 | 59.38 |
| 65 | 31 | 19 | 61.29 |
| 70 | 13 | 1 | 7.69 |
| 71 | 14 | 2 | 14.29 |
| 80 | 13 | 1 | 7.69 |
| 81 | 13 | 1 | 7.69 |
| 93 | 22 | 1 | 4.55 |
| 94 | 32 | 3 | 9.38 |
| 95 | 21 | 1 | 4.76 |

## V. Future Research Work

Our research work has proposed an agnostic model for streaming data life cycles in IoMT. Our experimental results show the potential of performing stream data management based on an end to end system architecture to leverage different resources, and make IoMT data available at real-time. Currently. our system supports one source of data streams, however, we plan to extend the edge layer to include sensors in order to monitor for example, weather conditions, humidity, air quality, and passenger ridership. Moreover, we would also like to explore the use of a temporary storage at the edge; firstly by defining where, how, and for how long to store the real-time and delayed IoMT data streams. Finally, we would like to announce that EFF has recently been integrated to Kinetic Cisco platform to fulfill overall IoMT solutions. Therefore, our future research work also considers to use, and migrate to this newest platform.

## Acknowledgments

This research was fully supported by the NSERC/CISCO Industrial Research Chair in Real-time Mobility Analytics. The authors are grateful to CODIAC Transit for providing the data streams used in this study, and Compute Canada for hosting one virtual machine that was used for the implementation of the cloud layer. Finally, we would like to thank Rimot for their support in fog node configuration.